\documentstyle[preprint,aps,floats,epsf]{revtex}
\tighten
\begin{document}
\draft
\title{Model for Polarized and Unpolarized Parton Density\\
Functions in the Nucleon}

\author{R.S.~Bhalerao$^a$,
N.G.~Kelkar$^b$,
and B.~Ram$^c$}

\address{  
$^a$Nuclear Theory Center, IUCF, Indiana University,\\
Bloomington, IN 47405, USA\\
and\\
Department of Theoretical Physics, 
Tata Institute of Fundamental Research,\\
Homi Bhabha Road, Colaba, Mumbai 400 005 INDIA\\
$^b$Nuclear Physics Division, 
Bhabha Atomic Research Center,\\
Trombay, Mumbai 400 085, INDIA\\
$^c$Department of Physics, 
New Mexico State University,\\
Las Cruces, New Mexico 88003-0001, USA}

\maketitle

\begin{abstract}
We present a physical model for polarized and unpolarized structure
functions and parton density functions (PDFs) of the proton and the
neutron. It reproduces the data on $F_2^p(x,Q^2)$ for $0.00001<x<1$
and $2.5<Q^2<5000$ GeV$^2$, $F_2^p(x) - F_2^n(x),~ F_2^n(x) /
F_2^p(x),~ xg(x),~ {\bar d}(x) - {\bar u}(x),~ d(x)/u(x)$, the
Gottfried sum, the fractional momentum of charged partons and the
polarized structure functions $g_1^{p,n}(x)$, at various $Q^2$. We
present for the first time, proton and neutron PDFs which do not
assume charge symmetry. Contrary to the common practice, we explain
polarized and unpolarized data with a single model.
\end{abstract}



\newpage
\section{Introduction}

Several new measurements of the polarized structure functions (SFs) of
the proton and the neutron have been reported in the last few years
\cite{g1pn}. In addition, recent experiments have significantly
widened the kinematic range over which the unpolarized SFs and gluon
density in the proton are known \cite{hera}. The New Muon
Collaboration has obtained accurate, final results on the ratio of the
deuteron and the proton (unpolarized) SFs, which can be used to
extract the ratio and the difference of the proton and the neutron SFs
\cite{nmc1}. There are new data also on the anti-down and anti-up
quarks in the proton \cite{dbub}. On the theoretical side, many
parameterizations of unpolarized and polarized parton density
functions (PDFs) exist \cite{para}. Parameter values are determined
from a global fit to the relevant high-energy data. As new data become
available, these parameterizations are generally revised.

In this Letter, we adopt a very different approach. Our input PDFs are
based on a different {\it ansatz} and have a few physically motivated
free parameters. In principle, two of these parameters ($a$ and $b$ in
(2)) could be evaluated theoretically, but in the absence of a full
understanding of a quantum chromodynamic (QCD) bound state, namely the
nucleon, we adopt the following pragmatic approach. We determine them
by fitting data on the unpolarized structure function $F_2(x,Q^2)$ at
{\it only one} value of $Q^2$. Here $x$ is the Bjorken variable and
$Q^2$ is the momentum scale. Thus we do not perform a global fit, with
a large number ($\sim$ 15-20) of free parameters in the PDFs. We do
not change the parameterization when we go from unpolarized to
polarized PDFs. This is contrary to the common practice where they are
parameterized separately. {\it Finally, in contrast to the commonly
available PDF sets, our PDFs for the proton and the neutron are not
based on the assumption of charge symmetry.} The issue of
charge-symmetry violation (CSV) has acquired importance in recent
months \cite{csv}.

Once the parameters of the model are determined, it can successfully
predict the rest of the unpolarized and polarized data, at various
values of $Q^2$ and $x$.
We have calculated:
$F_2^p(x,Q^2)$ for $0.00001<x<1$ and $2.5<Q^2<5000$ GeV$^2$, and the
following observables at the various values of $Q^2$ for which data
are available: $F_2^p(x) - F_2^n(x),~ F_2^n(x) / F_2^p(x),~ xg(x),
~{\bar d}(x) - {\bar u}(x),~ d(x)/u(x)$, 
the Gottfried sum $S_G$, the longitudinal
momentum fraction carried by quarks and antiquarks, and finally the
polarized structure functions $g_1^p(x)$ and $g_1^n(x)$. 
The agreement with experimental data is nearly as good as that
obtained by the standard PDFs \cite{para}.

\section{Model}

There is enough evidence to seriously consider the {\it ansatz} that
the input-scale parton densities in the nucleon may be
quasi-statistical in nature \cite{stat,rsb}. It is also known that
statistical mechanics is applicable to isolated quantum systems with
finite numbers of particles if the residual two-body interaction is
sufficiently strong and that the interaction-driven statistical
equilibrium that emerges in such systems can be described in terms of
the usual statistical quantities such as temperature \cite{flam}. We
do not claim that the formalism in \cite{flam} is applicable {\it in
toto} to the QCD bound state such as the nucleon. {\it We do, however,
believe that} \cite{stat,rsb,flam} {\it point towards the need for
having an open mind about the above ansatz.} The success of a
statistical model such as the present one, in reproducing and
correlating a vast body of polarized and unpolarized SF and PDF data
provides a strong a posteriori justification for the above {\it
ansatz}.

The parton number density $dn^i/dx$ in the infinite-momentum frame
(IMF) and the density $dn/dE$ in the nucleon rest frame are related to
each other by
\begin{equation}
{dn^i \over dx} 
= {M^2 x \over 2} \int ^{M/2}_{xM/2} {dE \over E^2}~~ {dn \over dE},
\eqnum{1} 
\end{equation}
where the superscript $i$ refers to the IMF, $M$ is the nucleon mass
and $E$ is the parton energy in the nucleon rest frame
\cite{rsb}. {\it This is a general relation connecting the two frames;
the only assumption made is that of massless partons.} This assumption
is common in the formalism of deep inelastic scattering.

There is a standard procedure in statistical mechanics to introduce
the effects of the finite size of an enclosure, in the expression for
the density of states \cite{hw}. It allows us to write $dn/dE$ for the
nucleon as
\begin{equation} 
dn/dE = g ~f(E)~ (VE^2/2\pi^2 + aR^2E + bR),
\eqnum{2}
\end{equation}
where $g$ is the spin-color degeneracy factor, $f(E)$ is the usual
Fermi or Bose distribution function $f(E) = \{\exp[(E-\mu)/T]\pm
1\}^{-1}$, $V$ is the nucleon volume and $R$ is the radius of a sphere
with volume $V$. The three terms in (2) are the volume, surface and
curvature terms, respectively; in the thermodynamic limit only the
first survives. We determined the parameters $a$ and $b$ in (2) by
fitting the structure function $F_2(x)$ data at a fixed momentum scale
$Q^2$. Unlike $a$ and $b$, the temperature $(T)$ and chemical
potential $(\mu)$ in the Fermi and Bose distributions are not fitted
to a detailed shape of any data, but get determined due to number and
momentum constraints on the PDFs (see below). Note also that $a,~b,~T$
and $\mu$ are constants independent of $x$ and $Q^2$.

We now depart from the procedure followed in \cite{rsb} and present a
more complete model for polarized as well as unpolarized PDFs and
SFs of protons and neutrons. If $n_{\alpha(\bar
\alpha)\uparrow(\downarrow)}$ denotes the number of quarks
(antiquarks) of flavor $\alpha$ and spin parallel (antiparallel) to
the nucleon spin, then any model of PDFs in the proton has to satisfy
the following seven constraints:
\begin{eqnarray}
n_{u\uparrow}+n_{u\downarrow}-n_{\bar u \uparrow}-n_{\bar u\downarrow}
&=&2, \eqnum{3}\\
n_{d\uparrow}+n_{d\downarrow}-n_{\bar d \uparrow}-n_{\bar d\downarrow}
&=&1, \eqnum{4}\\
n_{s\uparrow}+n_{s\downarrow}-n_{\bar s \uparrow}-n_{\bar s\downarrow}
&=&0, \eqnum{5}\\
n_{u\uparrow}-n_{u\downarrow}+n_{\bar u \uparrow}-n_{\bar u\downarrow}
&=& \Delta u, \eqnum{6}\\
n_{d\uparrow}-n_{d\downarrow}+n_{\bar d \uparrow}-n_{\bar d\downarrow}
&=& \Delta d, \eqnum{7}\\
n_{s\uparrow}-n_{s\downarrow}+n_{\bar s \uparrow}-n_{\bar s\downarrow}
&=& \Delta s, \eqnum{8}\\
\sum_{all~ partons} ({\rm momentum~~ fraction}) &=& 1, \eqnum{9}
\end{eqnarray}
and similarly for the neutron. The values of $\Delta u, ~\Delta d$ and
$\Delta s$ in (6)-(8) have been measured by several groups. We use
$\Delta u = 0.83 \pm 0.03, ~\Delta d = -0.43 \pm 0.03, ~\Delta s =
-0.10 \pm 0.03$ for the proton and $\Delta u = -0.40 \pm 0.04, ~\Delta
d = 0.86 \pm 0.04, ~\Delta s = -0.06 \pm 0.04$ for the neutron
\cite{g1pn}. The summation in (9) runs over quarks, antiquarks and
gluons. The numbers $n_{\alpha(\bar \alpha)\uparrow(\downarrow)}$ in
(3)-(8) are obtained from (1)-(2) by integrating the appropriate
$dn^i/dx$ over $x$, and the momentum fractions in (9) are obtained
similarly by integrating the appropriate $x dn^i/dx$ over $x$. The
double integrations are performed by first interchanging the order of
the two integrations, and then doing the $x$ integration analytically
and the $E$ integration numerically. It is necessary to distinguish
between $\mu_{\alpha \uparrow}$ and $\mu_{\alpha \downarrow}$, to
ensure that the statistical model is consistent with the constraints
(6)-(8). We note that $\mu_{\bar \alpha \uparrow} = - \mu_{\alpha
\downarrow}$ and $\mu_{\bar \alpha \downarrow} = -\mu_{\alpha
\uparrow}$. Hence (3)-(9) represent 7 coupled nonlinear equations in 7
unknowns, namely $\mu_{u \uparrow},~\mu_{u \downarrow},~\mu_{d
\uparrow},~\mu_{d \downarrow},~\mu_{s \uparrow},~\mu_{s \downarrow}$
and $T$. We can solve them numerically for a given choice of $a$ and
$b$. Once these 7 unknowns are determined, $dn^i/dx$ are known, and
polarized and unpolarized SFs can be evaluated using the standard
relations between SFs and PDFs.

We have described above our model of the {\it input-scale} ($Q_0^2$)
PDFs. Since data are available at various values of $Q^2$, it is
necessary to be able to evolve these PDFs from $Q_0^2$ to any other
$Q^2$. This involves evolving singlet, nonsinglet and gluon densities
in the unpolarized and polarized cases. We have done this by solving
the Dokshitzer-Gribov-Lipatov-Altarelli-Parisi equations in the
next-to-leading order, taking $Q_0^2 = M^2$, $N_f = 4$ and
$\alpha_s(m_z^2)=0.117$ \cite{pdg}.

\section{Results and Discussion}

Our fits to the structure functions $F_2^p(x)$ and $F_2^n(x)$ at $Q^2
= 4$ GeV$^2$ are shown in Fig. 1a. We chose 4 GeV$^2$, because the NMC
data \cite{nmc1} on $F_2^p-F_2^n$ are available only at this $Q^2$.
The resultant values of the two parameters, namely $a$ and $b$, are
given in Table I. We shall comment on these values later. It is easy
to show analytically using (3)-(8) that $\mu_{u \uparrow} > \mu_{d
\downarrow} > \mu_{u \downarrow} > \mu_{d \uparrow} > \mu_{s
\downarrow} = - \mu_{s \uparrow} > 0$, for the proton.  Figure 1b
shows our results for $F_2^p(x)-F_2^n(x)$ at $Q^2=4$ GeV$^2$ in
comparison with the NMC data. The discrepancy between the calculated
results and the data, near $x \simeq 10^{-2}$, is of the order of
0.005 which is negligible compared to $F_2^{p,n}$ in this region. The
calculated Gottfried sum at $Q^2 = 4$ GeV$^2$, over the interval
$0.004 < x < 0.8$, is 0.215 in agreement with the experimental number
$0.221 \pm 0.019({\rm syst}) \pm 0.008({\rm stat})$ \cite{gott}.

Once the parameters of the model are determined, it predicts the rest
of the unpolarized and polarized data quite successfully without any
further fitting. We now present these results. Figure 2 shows the
calculated $F_2^p$ in comparison with the data, in the range
$0.00001<x<1$ and $2.5<Q^2<5000$ GeV$^2$.

From SFs, we now turn to PDFs. The HERA and EMC data on $xg(x)$, the
Fermilab data on ${\bar d}(x) - {\bar u}(x)$ and the CDHSW data on
$d(x)/u(x)$ are compared, in Fig. 3, with our results and the results
based on parameterizations in \cite{para}. In particular, the
$d(x)/u(x)$ ratio in the limit $x \rightarrow 1$ is found to be 0.22
in good agreement with the QCD prediction 0.2 \cite{fj}. The momentum
fraction carried by quarks and antiquarks, at $Q^2= 15$ GeV$^2$ is
0.58 which is consistent with the MRST result \cite{para} and the
experimental observation that the charged partons carry about half the
proton momentum.

Finally, the results on the polarized structure functions $g_1^p(x)$
and $g_1^n(x)$ in Fig. 4, show that the model is able to predict the
shapes of the polarized data, once the parameters are determined as
explained in Sec. II. {\it The present model trivially satisfies the
general positivity constraints on the polarized ($\delta f$) and
unpolarized ($f$) PDFs}: $|\delta f(x,Q^2)| \leq f(x,Q^2)$.

We sum up the differences between the present model and that presented
in \cite{rsb}. (a) The earlier version did not distinguish between
$\mu_{\uparrow}$ and $\mu_{\downarrow}$. It involved solving three
simultaneous equations, namely the two number constraints and one
momentum constraint, for the three unknowns, namely $T, ~\mu_u$ and
$\mu_d$. It could not explain the polarized data. The present version
splits up each chemical potential into two. It involves solving the
seven simultaneous equations (3)-(9), and is able to explain and
correlate the polarized and the unpolarized data by means of a single
model. (b) We have presented here proton and neutron PDFs
without assuming charge symmetry. Whereas the subject of charge
symmetry violation in PDFs is being discussed extensively in the
literature \cite{csv}, we have presented PDFs which actually
incorporate this feature. (c) The present model is in agreement with
the observed difference and ratio of $F_2^p$ and $F_2^n$, which was
not the case earlier \cite{rsb}. (d) In \cite{rsb}, the QCD evolution
was performed to the leading order; here it is performed up to
next-to-leading order.

Interestingly, the signs as well as magnitudes of $a$ and $b$ (Table
I) turn out to be consistent with those in \cite{hw}, suggesting a
physical basis for our parameterization. Does the statistical model
provide a physical basis to the commonly used PDF parameterizations
\cite{para}? To investigate, we have compared our PDFs with those in
\cite{para}, at a common, low $Q^2=1.25$ GeV$^2$. Our $xu_v$ and
$xd_v$ have the same shapes and similar magnitudes as those in
\cite{para}. Our $xg$ is somewhat larger and $x \bar u$ and $x \bar d$
are somewhat smaller; however, their shapes are the same as in
\cite{para}. On this basis one would be inclined to answer the above
question in the affirmative. Details will be published elsewhere.

In conclusion, we have shown that the ideas from statistical mechanics
work even inside the nucleon. Trying to understand why they work,
could deepen our understanding of hadron structure as well as
statistical mechanics. Whereas all available PDFs assume charge
symmetry, this paper presents, for the first time, a set of proton and
neutron PDFs which does not make this simplifying assumption. The
model is remarkably successful in reproducing a large body of
polarized and unpolarized, PDF and SF data on the proton and the
neutron. Thus it is potentially able to make {\it quantitative}
predictions for the input PDFs. The scope and precision of the model
can be extended systematically --- e.g., by having a more elaborate
treatment of the finite-size effects, by allowing finite mass of the
charm quark, by considering nuclear corrections in the deuteron and
higher-twist effects, etc.


We are very grateful to A. Deshpande, D. Fasching and E.-M. Kabuss
for many useful communications.

\bigskip\bigskip\bigskip

Note added: Boros {\it et al.} \cite{csv} proposed a large CSV of the
sea quarks in the nucleon, as an explanation of the discrepancy
between neutrino (CCFR) and muon (NMC) nucleon structure function data
at low $x$. This, however, has been criticized by Bodek {\it et al.}
\cite{csv} who showed that the above proposal is ruled out by the
published CDF W charge asymmetry measurements. This controversy does
not affect the present work because the discrepancy, if any, between
neutrino and muon data at low $x$ is not used as an input anywhere in
the model. The discussion of this issue in the recent literature was
used only as one of the motivations for this model. The origin of CSV
in the present model (i.e. the origin of the different values of $a$
and $b$ for the proton and the neutron) is in the tiny differences
between $\Delta u, \Delta d, \Delta s$ values for the proton and the
neutron (see the RHSs of Eqs. (6)-(8)), and also in the fact that we
fit to $F_2^p$ and $F_2^n$ separately. These two points are
independent of the above controversy.




\begin{table}[tb]

\caption{Values of the parameters $a$ and $b$. Temperature ($T$)
and chemical potentials ($\mu$) are in MeV. $T$ and $\mu$
are constrained by (3)-(9) once $a$ and $b$ are
specified.}

\label{tab:1}

\begin{tabular}{ccc}
                    & Proton  & Neutron \\ \hline
$(a,b)$             &$(-0.376, 0.504)$   &$(-0.300, 0.504)$ \\ \hline 
$T$                                      & $62$        & $59$       \\
$(\mu_{u\uparrow}, \mu_{u\downarrow})$   & $(210,86)$  & $(40,94)$  \\
$(\mu_{d\uparrow}, \mu_{d\downarrow})$   & $(42,106)$  & $(188,76)$ \\
$(\mu_{s\uparrow}, \mu_{s\downarrow})$   & $(-7,7)$    & $(-4,4)$   \\
\end{tabular}
\end{table}


\newpage
\begin{figure}
\vspace{10cm}
\includegraphics{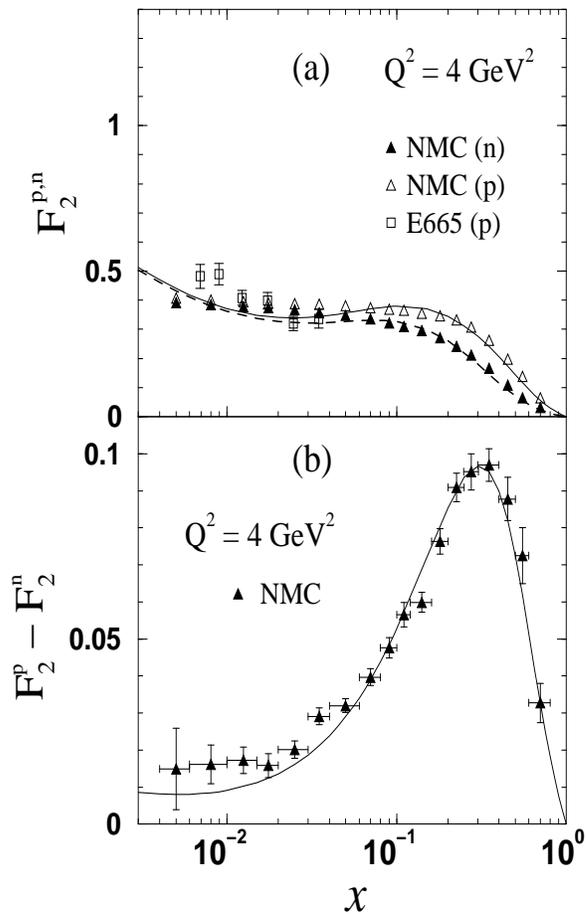}
\caption
{(a) $F_2^p$ and $F_2^n$ {\it vs.} $x$. The solid and dashed curves
are our fits to $F_2^p$ and $F_2^n$ data, respectively. The $F_2^p$
data are from [12,13] and the $F_2^n$ data are extracted from the
accurate final results on $F_2^p - F_2^n$ at $Q^2 = 4$ GeV$^2$,
supplied by the NMC collaboration [3]. NMC obtained these results
by ignoring nuclear corrections in the deuteron.
Since at low $x$, neutron data
are not available, the proton data were used in that region, for the
purpose of the fit. (b) $F_2^p - F_2^n$ {\it vs.} $x$. The curve
represents our calculation and the data are from [3]. The small
discrepancy between the curve and the data at low $x$,
is discussed in the text.}  
\end{figure}


\newpage\newpage
\begin{figure}
\vspace{20cm}
\includegraphics{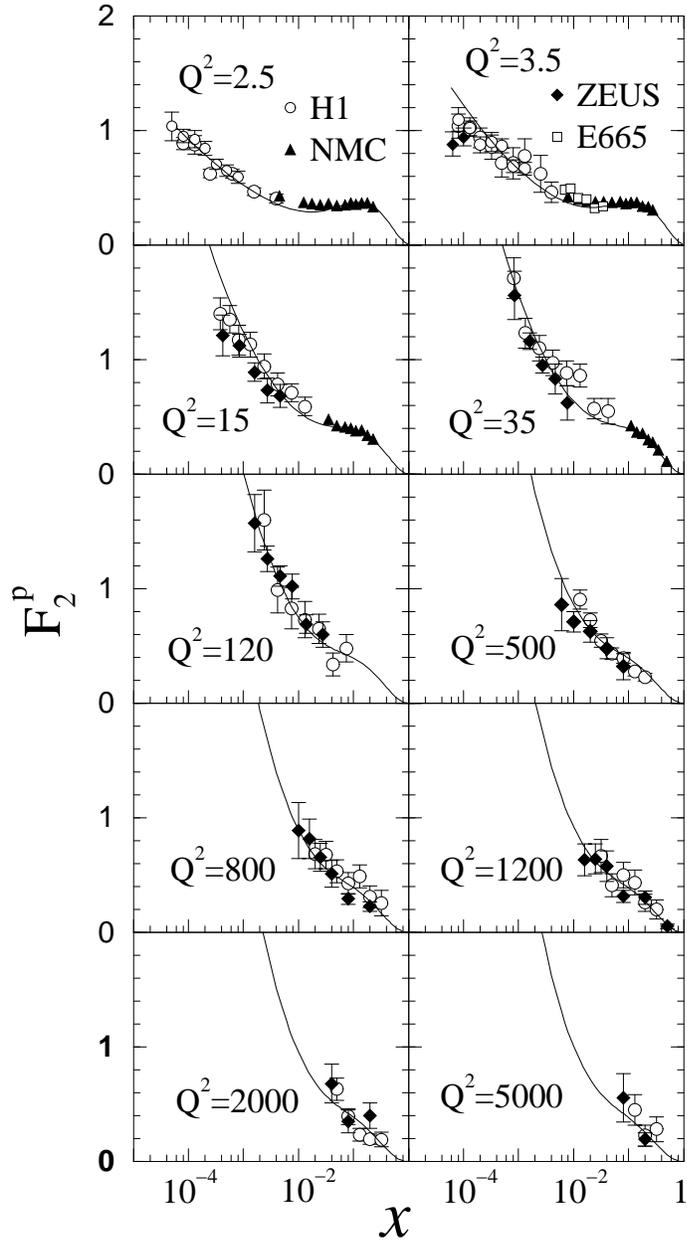}
\caption{Predictions of the model, for the unpolarized structure
function $F_2^p$ {\it vs.} $x$, over a wide range of $Q^2$ (GeV$^2$)
and $x$, compared with experimental data [2,12,13].}
\end{figure}


\newpage
\begin{figure}
\vspace{20cm}
\includegraphics{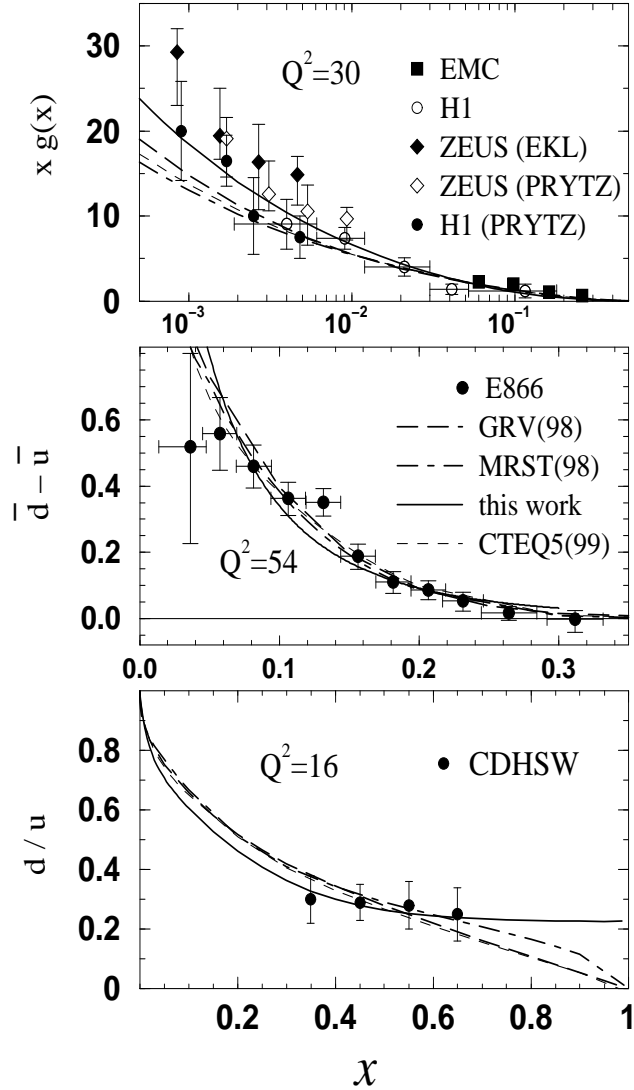}
\caption{Predictions of the model, for parton density functions. Gluon
data from [16], $(\bar d - \bar u)$ data from [4] and $d/u$ data from
[17]. In the top panel, data are at 20 and 30 GeV$^2$ while all the
calculated results are at 30 GeV$^2$.}
\end{figure}


\newpage
\begin{figure}
\vspace{20cm}
\includegraphics{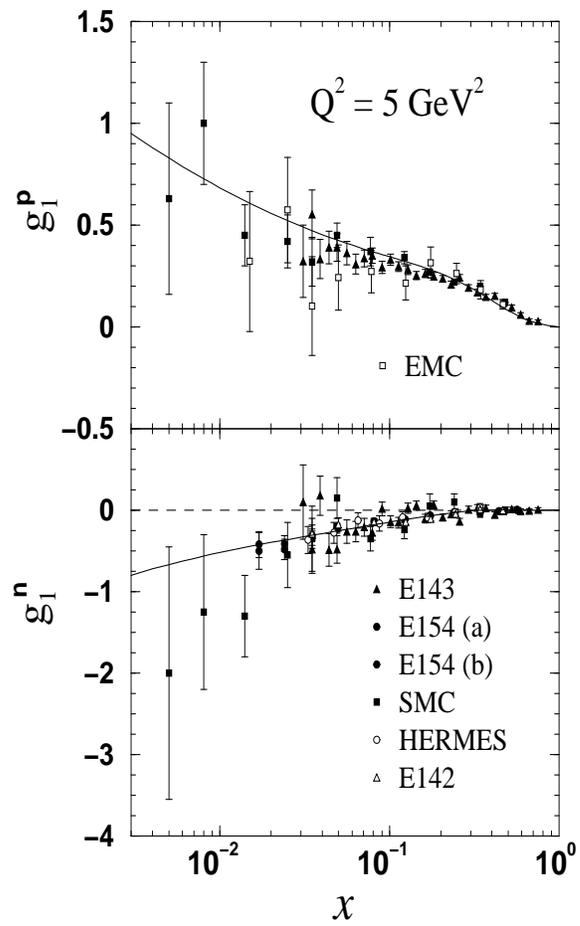}
\caption{Predictions of the model, for polarized SFs. Data are from [1].}
\end{figure}

\end{document}